\documentclass[sigconf,nonacm,dvipdfmx]{acmart}
\usepackage[labelformat=parens, subrefformat=parens]{subcaption}

\AtBeginDocument{%
  \providecommand\BibTeX{{%
    \normalfont B\kern-0.5em{\scshape i\kern-0.25em b}\kern-0.8em\TeX}}}

\begin{document}

\title{Toward a view-based data cleaning architecture}

\author{Toshiyuki Shimizu}
\affiliation{%
  \institution{Graduate School of Informatics, Kyoto University}
  \city{Kyoto}
  \country{Japan}
  \postcode{606-8501}
}
\email{tshimizu@i.kyoto-u.ac.jp}

\author{Hiroki Omori}
\affiliation{%
  \institution{Graduate School of Informatics, Kyoto University}
  \city{Kyoto}
  \country{Japan}
  \postcode{606-8501}
}
\email{hiroki@db.soc.i.kyoto-u.ac.jp}

\author{Masatoshi Yoshikawa}
\affiliation{%
  \institution{Graduate School of Informatics, Kyoto University}
  \city{Kyoto}
  \country{Japan}
  \postcode{606-8501}
}
\email{yoshikawa@i.kyoto-u.ac.jp}

\begin{abstract}
Big data analysis has become an active area of study 
with the growth of machine learning techniques. 
To properly analyze data, 
it is important to maintain high-quality data. 
Thus, research on data cleaning is also important. 
It is difficult to automatically detect and correct inconsistent values 
for data requiring expert knowledge or 
data created by many contributors, such as 
integrated data from heterogeneous data sources. 
An example of such data is metadata for scientific datasets, 
which should be confirmed by data managers while handling the data. 
To support the efficient cleaning of data by data managers, 
we propose a data cleaning architecture 
in which data managers interactively browse and correct portions of data through views. 
In this paper, we explain our view-based data cleaning architecture and 
discuss some remaining issues. 
\end{abstract}



\maketitle

\section{Introduction}
\label{sec:intro}

The use of big data is attracting attention, and 
research on machine learning is very active. 
To properly analyze data, 
it is important to maintain high-quality data. 
Therefore, data wrangling~\cite{wrangling} is an emerging research topic, 
as the process of data exploration and transformation is essential. 
A considerable amount of effort is required to make data usable, 
and the majority of this effort is spent preparing data~\cite{exploratory}.

In this paper, we focus on data cleaning and 
propose a data cleaning architecture in which data managers interactively manipulate data. 
Interactive data cleaning is especially effective 
for data that are difficult to automatically clean by static rules~\cite{continuous} or 
by machine learning. 
One example of such data is scientific metadata, 
which describe details 
such as dataset name, category, and data provider, 
from scientific datasets. 
Scientific metadata convey useful information about the target dataset, and each dataset is
described by a person in charge of the dataset, who is a contributor to the data. 
Therefore, the metadata tend to contain inconsistent values 
depending on the data contributor.  
Note that although using dictionaries to manage the terms in metadata will improve the situation, 
developing dictionaries in advance is difficult for scientific metadata, 
as their production requires expert knowledge 
and new words, which may belong to only a particular research field, 
need to be continuously updated.

Figure \ref{table1} shows a toy example of scientific metadata. 
Each tuple corresponds to metadata from the dataset.  
ID stands for Identifier, 
NP stands for Name of Provider, 
OP stands for Organization of Provider, 
OC stands for Organization for Contact, and 
Y stands for Year.
In this example, the values in the NP column are inconsistent, 
as the order of the family name and given name varies. 
Additionally, the value of `KU' in the OP column is ambiguous, as we cannot easily determine 
whether it means `Kyoto Univ.' or `Kobe Univ.'
We require background knowledge to handle these values,  
making them difficult to automatically correct.

\begin{figure}[t]
  \begin{center}
    \begin{tabular}{lllllll}
      \toprule
      ID & NP & OP & OC & Y \\
      \midrule
      1  & Omori Hiroki & KU & Yoshikawa Lab. & 2011 \\
      2  & Hiroki OMORI & Kyoto Univ. & Yoshikawa Lab. & 2016 \\
      3  & Taro YAMADA & Kobe Univ. & Suzuki Lab. & 2017 \\
      4  & Hiroki OMORI & Kobe Univ. & Suzuki Lab. & 2018 \\
      \bottomrule
    \end{tabular}
  \caption{An example of unclean data.}\label{table1}
\end{center}
\end{figure}

To deal with inconsistencies in values that are difficult to automatically correct, 
it is essential for data managers to confirm data. We 
propose a data cleaning architecture to support data managers in conducting data cleaning 
while they browse and evaluate data.

We assume target data to be cleaned are stored in relational databases and 
utilize {\it view} to support data cleaning. 
The target users of our system are data managers. 
Users interactively browse and evaluate data 
through three operations: data marking, view creation, and data correction. 
By selecting a portion of data using a view, 
users can focus on only the portion of interest, and 
the system can support efficient data cleaning.

In this paper, we explain the overview of our view-based data cleaning architecture.  
We also discuss some remaining issues in the architecture, 
the improvement of which would better support the users. 
The remainder of this paper is organized as follows:
Section \ref{sec:related} introduces related work. 
Section \ref{sec:architecture} explains our architecture, and 
a case study using actual scientific metadata is shown in Section \ref{sec:case}. 
We discuss some issues in Section \ref{sec:discussions} 
and conclude the paper in Section \ref{sec:conclusion}.

\section{Related Work}
\label{sec:related}

We adopt interactive data cleaning, 
the features of which are discussed in the work by Volkovs et al.~\cite{continuous}. 
The target databases of traditional data cleaning are ones with static integrity constraints. 
However, we are more curious about dynamic environments.
Because we need to care about updating integrity constraints, 
batch processing has limitations for dynamic environments.

Falcon~\cite{falcon} and ICARUS~\cite{icarus} are interactive data cleaning systems. 
Falcon~\cite{falcon} constructs update queries from repairs, and users confirm the update queries 
for correcting data. 
ICARUS~\cite{icarus}, whose architecture is similar to ours, 
shows a subset of data to users for data cleaning. 
The purpose of ICARUS is to fill potentially inappropriate null data values. 
We consider the introduction of the concept of views to interactive data cleaning 
to be the novelty of our architecture.

Recently, research on data cleaning using machine learning techniques has been active. 
HoloDetect~\cite{holodetect} is an error detection method that uses machine learning. 
HoloDetect uses supervised machine learning and applies few-shot learning to error detection. 
Although it is a very powerful method, 
it assumes the target data are in the Probabilistic Unclean Database~\cite{pud} 
and that the main purpose is {\it error} detection. 
We consider that data like scientific metadata, 
which require expert knowledge and are created by many contributors, 
are hard to handle in the probabilistic manner 
as they often contain {\it inconsistent} values depending on the contributor. 
Data managers need to be involved in order to handle inconsistent values.

\section{Our Architecture}
\label{sec:architecture}

We propose a data cleaning architecture in which users interactively correct unclean data
stored in relational databases. 
Our target unclean data contain inconsistent values.

\begin{figure*}[t]
  \begin{tabular}{cc}
  \begin{minipage}{9cm}
    \begin{center}
    \begin{tabular}{lllllll}
      \toprule
      ID & NP & OP & OC & Y \\
      \midrule
      1  & OMORI & KU & Yoshikawa Lab. & 2018 \\
      2  & OMORI & KU & Yoshikawa Lab. & 2019 \\
      3  & YAMADA & kyoto univ. & kyoto univ. & 2016 \\
      4  & SUZUKI & Kyoto Univ. & Kyoto Univ. & 2016 \\
      5  & SUZUKI & Kyoto Univ. & Kyoto Univ. & 2014 \\
      ... & ... & ... & ... & ... & ... \\
      \bottomrule
    \end{tabular}
  \end{center}
  \subcaption{Original data}
  \label{table2}
  \end{minipage}
  &
  \begin{minipage}{9cm}
    \begin{center}
    \begin{tabular}{lllllll}
      \toprule
      ID & NP & OP & OC & Y \\
      \midrule
      1  & OMORI & KU & Yoshikawa Lab. & 2018 \\
      2  & OMORI & KU & Yoshikawa Lab. & 2019 \\
      8  & OMORI & KU & ylab & 2016 \\
      12 & KATO & KU & Tanaka Lab. & 2013 \\
      13 & KATO & KU & Tanaka Lab. & 2014 \\
      ... & ... & ... & ... & ... & ... \\
      \bottomrule
    \end{tabular}
    \subcaption{View: OP=`KU'}
    \label{table3}
  \end{center}
  \end{minipage}
  \\ \\
  \begin{minipage}{9cm}
    \begin{center}
  \begin{tabular}{lllllll}
    \toprule
      ID & NP & OP & OC & Y \\
    \midrule
    1  & OMORI & KU & Yoshikawa Lab. & 2018 \\
    2  & OMORI & KU & Yoshikawa Lab. & 2019 \\
    8  & OMORI & KU & ylab & 2016 \\
    34 & OMORI & KU & ylab & 2016 \\
    49 & OMORI & KU & ylab & 2015 \\
    ... & ... & ... & ... & ... & ... \\
    \bottomrule
  \end{tabular}
  \subcaption{View: OP=`KU' AND NP=`OMORI'}
  \label{table4}
\end{center}
\end{minipage}
 &
\begin{minipage}{9cm}
  \begin{center}
  \begin{tabular}{lllllll}
    \toprule
      ID & NP & OP & OC & Y \\
    \midrule
    1  & OMORI & KU & Yoshikawa Lab. & 2018 \\
    2  & OMORI & KU & Yoshikawa Lab. & 2019 \\
    8  & OMORI & KU & ylab & 2016 \\
    20 & OMORI & kyoto-u & ylab & 2013 \\
    21 & OMORI & kyoto-u & ylab & 2014 \\
    ... & ... & ... & ... & ... & ...  \\
    \bottomrule
  \end{tabular}
  \subcaption{View: NP=`OMORI'}
  \label{table5}
  \end{center}
\end{minipage}
\end{tabular}
\caption{Examples of original data and the views.}
\label{tables}
\end{figure*}

\subsection{User Operations}
\label{sec:operations}

We considered three operations: data marking, view creation, and data correction. 
By using these three operations iteratively, users can efficiently browse and correct data.

\begin{enumerate}
  \item Data marking: Operation that marks values that users consider unclean. 
  \item View creation: Operation that creates a view based on the marked values. 
  \item Data correction: Operation that corrects values within the view. 
\end{enumerate}

The basic flow of operations is as follows: 
users mark values, create a view based on the marked values, and correct values within the view. 
We explain the operations using the examples shown in Figure \ref{tables}. 
Figure \ref{tables} shows examples of data and the views imitating simple scientific metadata. 
Each tuple explains a corresponding dataset, 
and the table schema is the same as that of Figure \ref{table1}.

Figure \ref{table2} shows an original unclean dataset.  
We assume the `KU' values in OP in the first two tuples should actually be `Kyoto Univ.' 
In this situation, users will first mark two `KU's in Figure \ref{table2}. 
It is natural that users would like to focus on and browse tuples 
whose OP is `KU'. 
Therefore, users create a view with the condition OP=`KU', 
which is shown in Figure \ref{table3}. 
Users can confirm and correct values efficiently in this view, 
including other `KU's that were not marked in the first step. 
Note that our architecture is not intended to automatically correct data 
but to support the users in finding the inconsistent data. 
In the example in Figure \ref{table3}, 
users should consider and judge each `KU' separately,
as there may be `KU's that do not correspond to 'Kyoto Univ.'

\subsection{Iteration of the Operations}

Because we may have different kinds of inconsistent values 
in different portions of the dataset, 
users can iteratively repeat the operations we explained in Section \ref{sec:operations}.

Of particular importance are the cases in which 
users find other kinds of inconsistent values which were not marked in the first data marking step 
while users browse data in the created view. 
For example, users may want to mark `ylab' in the view in Figure \ref{table3}. 
When inconsistent values are found in the view step, 
it is possible to utilize the conditions of that view 
in order to create a new view for the newly found inconsistent values. 
In the running example, if users want to confirm the tuples of NP=`OMORI' 
in the view of Figure \ref{table3}, 
a new view with the conditions of OP=`KU' AND NP=`OMORI' (Figure \ref{table4}) can be created. 
The information on the conditions used to create new views 
can be helpful for finding other inconsistent values. 
Thus, it is also possible to further create a view 
with the condition NP=`OMORI' only (Figure \ref{table5}),
with the expectation that other inconsistent values can be found.

\section{Case Study}
\label{sec:case}

We observed actual scientific metadata managed by 
DIAS (Data Integration and Analysis System)\footnote{https://www.diasjp.net/en/} and 
applied our architecture to it. 
DIAS manages a variety of earth science datasets with metadata. 
As the original metadata are in XML format, we converted them into a relational format. 
For this observation, we use 426 datasets and focus on 12 attributes, 
which define the core of the information, 
such as dataset name, category, date of dataset, name of metadata author, and name of provider, 
for each dataset.

\begin{figure*}[t]
  \begin{tabular}{c}
  \begin{minipage}{8cm}
    \begin{center}
      \begin{tabular}{ll}
        \toprule
        OC & NA \\
        \midrule
        {\it JAMSTEC/DrC} & Hiromichi Igarashi \\
        {\it JAMSTEC/DrC} & Hiromichi Igarashi \\
        Center for Global Envir... & Shin-ichiro Nakaoka \\
        JAMSTEC/RIGC & Hiromichi Igarashi \\
        NULL & Japan coast guard hydrogr... \\
        Atmoshpere and Ocean... & Sachihiko Itoh \\
        Japan Agency for Marin... & JAMSTEC-CEIST \\
        Center for Global Envir... & Nojiri, Yukihiro \\
        Center for Global Envir... & Nojiri, Yukihiro \\
        {\bf DrC/JAMSTEC} & Hiromichi Igarashi \\
        ... & ... \\
        \bottomrule
      \end{tabular}
      \caption{A part of data in the actual scientific metadata.}\label{fig2}
    \end{center}
  \end{minipage}\\
  \\
  \begin{minipage}{15cm}
    \begin{center}
      \begin{tabular}{llll}
        \toprule
        NA & OA & NP & OP \\
        \midrule
        Hiromichi Igarashi & JAMSTEC/DrC & Sugiura, Nozomi, Dr. & {\bf JAMSTEC/DRC}\\
        Hiromichi Igarashi & JAMSTEC/DrC & Dr. Nozomi Sugiura & {\bf JAMSTEC/DRC}\\
        Hiromichi Igarashi & {\bf Japan Agency for Marine-Earth...} & Kazuo Umezawa & Japan Aerospace Exploration...\\
        Hiromichi Igarashi & {\bf Japan Agency for Marine-Earth...} & Kazuo Umezawa & Japan Aerospace Exploration...\\
        Hiromichi Igarashi & {\bf Japan Agency for Marine-Earth...} & Kazuo Umezawa & Japan Aerospace Exploration...\\
        Hiromichi Igarashi & {\bf Japan Agency for Marine-Earth...} & Kazuo Umezawa & Japan Aerospace Exploration...\\
        Hiromichi Igarashi & {\bf Japan Agency for Marine-Earth...} & Kazuo Umezawa & Japan Aerospace Exploration...\\
        Hiromichi Igarashi & JAMSTEC/DrC & Hiromichi Igarashi & JAMSTEC/DrC\\
        Hiromichi Igarashi & JAMSTEC/DrC & Hiroshi Kawamura & Center for Atmospheric and...\\
        Hiromichi Igarashi & JAMSTEC/DrC & Sugiura, Nozomi, Dr. & {\bf JAMSTEC/DRC}\\
        Hiromichi Igarashi & {\bf DrC/JAMSTEC} & Remote Sensing Systems & NULL\\
        ... & ... & ... & ...\\
        \bottomrule
      \end{tabular}
      \caption{A view of the actual data.}\label{fig3}
    \end{center}
  \end{minipage}
\end{tabular}
\end{figure*}

Figure \ref{fig2} shows a part of the metadata focusing on the attributes of 
OC, which stands for Organization for Contact, and 
NA, which stands for Name of the metadata Author. 
In Figure \ref{fig2}, we find `JAMSTEC/DrC', which is shown in italics, and 
`DrC/JAMSTEC', which is shown in boldface type. 
Because these values can be considered to indicate the same organization, 
these values are marked in this example. 
Figure \ref{fig3} shows a view with the condition NA=`Hiromichi Igarashi',
which we assume is the condition that covers the marked values. 
NA stands for the Name of the metadata Author, 
OA stands for the Organization of the metadata Author, 
NP stands for the Name of the Provider, and 
OP stands for the Organization of the Provider. 
In the view, we observe the values `JAMSTEC/DrC', `DrC/JAMSTEC', and 
`Japan Agency for Marine-Earth Science and Technology', which is the full name of JAMSTEC, in OA. 
Additionally, we observe `JAMSTEC/DRC', 
which is considered to be another variant of `JAMSTEC/DrC' in OP. 
By browsing data through views, users can focus on the interesting parts of the data 
and can consider strategies for efficient data correction. 
It is also expected that users can more easily find inconsistent values in the views.

\section{Discussion}
\label{sec:discussions}

In order to make our view-based data cleaning architecture more effective, 
we are planning to develop support functions,
especially for data marking and view creation.

\subsection{Support for Data Marking}

There are two main approaches for the support of data marking: 
one is the rule-based approach, and the other is the machine learning approach. 
As we noted in Section \ref{sec:related}, 
we think error detection methods using machine learning are difficult to apply 
to data like scientific metadata, which are constructed by many contributors. 
As for the rule-based approach, 
there is some research that utilizes the concept of functional dependencies (FDs)
in order to find inconsistent values~\cite{cleaning1, cleaning2, cleaning3}. 
The methods using FDs are based on the idea of 
finding a minimum set of tuples to be removed 
to hold the FDs~\cite{TANE, fastFD}. 
This approach is effective for simple data; however, it has limitations for complex data, 
such as the case in which there are multiple affiliations for one personnel.

Conditional functional dependencies (CFDs)~\cite{CFDfirst, DiscoveringCFD} are 
an extension of the concept of functional dependencies. 
CFDs consider dependencies on pattern tuples 
and can flexibly define constraints among values. 
CFDs are also used for data cleaning~\cite{findCFDtableau,semandaq,holoclean}.

Because data like scientific metadata are created by many contributors, 
we need to handle variation in values more carefully. 
The values in the subset of data that are input by the same data contributor may be consistent; 
however, they may be inconsistent compared to the rest of the dataset. 
We believe distributed representations of values~\cite{dr} 
can be effectively applied to our situation.

\subsection{Support for View Creation}

Creating proper views to target marked values is a challenging problem. 
To support view creation, 
we would like to create reasonable views automatically 
by considering common properties among the marked values. 
For example, if the values of another attribute in the tuples of the marked values are the same, 
we can utilize this shared value to construct queries for creating views. 
Tran et al. proposed constructing a query $Q'$ given the output of another query $Q$~\cite{qbo}, 
that is, constructing a condition for a given part of data. 
Proper views within our architecture have the following features
and require further discussion.

\begin{itemize}
\item For confirmation and comparison, the views contain values other than the marked values. Both tuples and attributes need to be considered for this point. 
\item The size of a view is moderate and allows users to browse. 
\end{itemize}

\section{Conclusion}
\label{sec:conclusion}

In this paper, we proposed a view-based data cleaning architecture 
that supports data managers in cleaning data. 
We considered that data cleaning for data like scientific metadata 
needs to be confirmed by data managers, and introduced three operations: 
data marking, view creation, and data correction. 
Additionally, we applied our architecture to actual scientific metadata and observed the results. 
We believe that our view-based architecture can provide some insights for data managers 
that are difficult to obtain using automatic solutions.

We would like to develop support functions for our architecture, 
as we discussed in Section \ref{sec:discussions}. 
We will also study the utilization of the knowledge acquired during data cleaning. 
We believe the recorded corrections can be utilized 
to predict desirable data cleaning~\cite{guided}.

\begin{acks}
This work was partially supported by JSPS KAKENHI Grant Numbers JP17H06099, JP18H04093, and JP18K11315.
\end{acks}

\bibliographystyle{ACM-Reference-Format}
\bibliography{references}



\end{document}